%% file: paper_la.tex
\documentclass[twocolumn,showpacs,aps,prl,superscriptaddress]{revtex4-1}
\usepackage{graphicx}
\usepackage{dcolumn}
\usepackage{amsmath}
\usepackage{epsfig}
\usepackage{colordvi}
\usepackage{color}
\usepackage{hhline}
\usepackage{multirow}
\usepackage{subfigure}
\usepackage{rotating}
\usepackage{hyperref}
\usepackage{tabularx}
\RequirePackage{xspace}

\def\pep2{PEP-II}

\usepackage{relsize}
\def\babar{\mbox{\slshape B\kern-0.1em{\smaller A}\kern-0.1em
    B\kern-0.1em{\smaller A\kern-0.2em R}}}

\newcommand {\sigwind}[0]{$5.27 < m_{\mathrm{ES}}<5.29$ \gevcc}

\newcommand {\sideband}[0]{ $m_{\mathrm{ES}}<5.27$ \gevcc }
\newcommand{\ups}{$\Upsilon (4S)$}
\newcommand{\fourp} {$ p p \bar{p} \bar{p}$}

\newcommand {\grunberg} {$\Bzb \rightarrow \Lambda^{+}_{c} p \bar{p} \bar{p}$}
\newcommand {\Mes} {$m_{\mathrm{ES}}$}
\newcommand {\bpairs} {$N_{{B\Bb}}$}
\newcommand{\fitrange}[0]{$5.2 < m_{\mathrm{ES}} <  5.3 $~\gevcc}
\newcommand {\significance} { $S/\sqrt{S+B}$}
\newcommand {\mymode}{$B^0\rightarrow$~\fourp}

\newcommand {\deltae}{$\Delta E$}

\newcommand {\pppipi}{$ p\bar{p} \pi \pi$}
\newcommand {\ppkk}{$ p\bar{p} KK$}
\newcommand {\ppkpi}{$ p\bar{p}K\pi$}
\newcommand{\BABARPubYear}    {17}
\newcommand{\BABARPubNumber}  {004}
\newcommand{\SLACPubNumber}   {17239}

\usepackage{relsize}
\def\babar{\mbox{\slshape B\kern-0.1em{\smaller A}\kern-0.1em
    B\kern-0.1em{\smaller A\kern-0.2em R}}}

\long\def\inst#1{\par\nobreak\kern 4pt\nobreak
    {\it #1}\par\vskip 10pt plus 3pt minus 3pt}

\input babarsym
\graphicspath{./}

\begin{document}

\begin{flushleft}
\vspace{-2cm}
%\babar\ Analysis Document \# 2701, Version 15\\
SLAC-PUB-\SLACPubNumber \\
\babar-PUB-\BABARPubYear/\BABARPubNumber
%\date{\today}
\end{flushleft}

\title{
\Large \bf \boldmath Search for the decay mode \mymode
} 
\input authors_mar2018_frozen.tex

\begin{abstract}
\noindent
A search is presented for the four-body decay \mymode\ in a sample of $471$ million $B\bar{B}$ pairs collected with the \babar\ detector, operated at the SLAC PEP-II asymmetric-energy \epem\ collider. The center-of-mass energy is $10.58$~\gev. From a fit to the distribution of the energy-substituted mass \Mes, the result $\mathcal{B}(\text{\mymode})=(1.1 \pm 0.5 \pm 0.2) \times 10^{-7}$ is extracted, where the first uncertainty is statistical and the second is systematic. The significance of the signal is $2.9$ standard deviations. The upper limit on the branching fraction is determined to be  $2.0 \times 10^{-7}$ at $90\%$ confidence level.

\end{abstract}

\pacs{13.25.Hw, 13.60.Rj, 14.40.Nd}% 

\maketitle
\noindent
The inclusive branching fraction of $B$~mesons decaying into final states with at least one baryon-antibaryon pair is approximately $7\%$~\cite{ref:inclusiveArgus}, while the sum of all measurements of exclusive baryonic channels is less than $1\%$~\cite{ref:pdg}. Recent measurements from the LHCb experiment~\cite{ref:LHCbBF, ref:pure_baryon, ref:bsubs} have raised new interest in this field.
Studying exclusive baryonic decays of $B$~mesons provides a deeper insight into the mechanism of hadronization into baryons and may allow a better understanding of the threshold enhancement effect, which is a dynamical enhancement, relative to the pure phase space expectation, of the production rate of baryon-antibaryon pairs at their invariant mass threshold. So far, this process is only qualitatively understood. Theoretical models, e.g., the quantum chromodynamics (QCD) sum rule~\cite{ref:qcdsum} and the perturbative QCD approach~\cite{ref:pqcd}, need validation and input from experimental data. Although the threshold effect is also observed in $B$ decays to charmed baryons~\cite{ref:threshold_charmed}, its effect is not as pronounced as in charmless three-body baryonic decays, where the peak at the threshold of the invariant baryon-antibaryon mass distribution was first observed~\cite{ref:threshold_charmless, ref:threshold_discover}. This enhancement could explain the hierarchy trend of the branching fractions for baryonic \B decays. It has been observed that many three-body final states have larger rates than their two-body counterparts, and also that some three-body decays are suppressed compared to the four-body case~\cite{ref:path_rare, ref:charmless_mult, ref:threshold_mult}.
The phenomenological approaches describe these observations in terms of gluonic and fragmentation mechanisms~\cite{ref:threshold_theory1} and pole models~\cite{ref:threshold_theory2}. For final states with a $p\bar{p}$ pair, a threshold enhancement could possibly arise from an intermediate X(1835) baryonium resonance, as proposed in Ref.~\cite{ref:baryonium}.

In this paper, we report on a search for \mymode\ decays (the inclusion of charge conjugate processes is implied throughout). The data were collected with the \babar\ detector~\cite{ref:detector_old, ref:detector} at the SLAC PEP-II asymmetric-energy $e^+e^-$ collider. The decay of a \B meson to two baryon-antibaryon pairs has not yet been observed. No quantitative predictions for this process are yet available. The measurement of a four baryon decay mode would provide useful information to help discriminate between existing models and aid in the development of phenomenological models for four-baryon production.
Previously, we performed a search for \grunberg\ decays, setting a 90\% confidence level (C.L.) upper limit on the decay branching fraction of $2.8\times 10^{-6}$~\cite{ref:grunberg}. Based on this result, using a scaling factor to account for the Cabibbo suppression for the $b \to u$ decay, and also taking into account the larger phase space of the final state, we estimate the branching fraction for the \mymode\ decay mode to be on the order of $10^{-7}$. We use this assumption to optimize the selection criteria. The threshold effect has been found to be enhanced for large values of $q$, which is the available momentum in the rest frame of the decaying \B, with no visible enhancement for $q$ values below about $200$~MeV$/c$~\cite{ref:bfactories}. This feature could explain the absence of an observed signal in \grunberg\ decays and at the same time could enhance the branching fraction for \mymode. Moreover, the \mymode\ decay rate may benefit from the low-invariant-mass enhancement of the double $p\bar{p}$ system and from presence of nontrivial intermediate bound states~\cite{ref:pp_paper}.

The analysis is based on the full data set collected with \babar\ at center-of-mass energy 10.58\,GeV, corresponding to the peak of the \ups\ resonance. The event sample contains $N_{B\Bb}=471\times10^6$ $B\Bb$ pairs, corresponding to integrated luminosity of $424$~fb$^{-1}$~\cite{ref:lumi_paper}. Charged-particle momenta are measured by means of a five-layer double-sided silicon vertex tracker and a 40-layer multiwire drift chamber, both operating in the $1.5$~T magnetic field of a superconducting solenoid. The particle identification (PID) for protons, kaons, and pions uses the specific energy loss measured in the tracking devices and the measurement of the Cherenkov angle provided by the internally reflecting, ring-imaging Cherenkov detector. We use Monte Carlo (MC) simulated events of the processes $e^+e^- \to B\bar{B}$, where the $B$~mesons decay generically according to known branching fractions and decay amplitudes~\cite{ref:pdg}, and $e^+e^- \to q \bar{q}$ (with $q=u,d,s,c$) to model the background. These samples correspond to at least three times the integrated luminosity of the data. In addition, we generate a sample of $687\,000$ signal decays $e^+e^- \to B^0\Bzb$, where one of the $B$~mesons decays into \fourp\ (referred to as the signal MC sample). Monte Carlo events are simulated with the \textsc{EvtGen} and \textsc{Jetset}~\cite{ref:bkg_mc, ref:continuum_mc} event generators, with the response of the detector simulated using the \textsc{Geant4} suite of programs~\cite{ref:geant}. Signal and background MC samples are used for the signal efficiency determination and for the modeling of the signal and background distributions.

A $B$ meson candidate is reconstructed by combining four charged tracks, two identified as protons and two as antiprotons, kinematically fitting them to a common vertex and requiring the fit probability to exceed $0.1\%$. The direction of the reconstructed \B meson is required to originate from the interaction region, which is constrained to the beam-spot size in the laboratory frame. Tracks are rejected if the combination of two oppositely charged tracks is found to be consistent with $K^0_S$ or $\Lambda$ hypotheses. 

Loose preselection requirements are applied to the kinematic variables~\cite{ref:bfactories} \Mes$=\sqrt{(E_{\mathrm{beam}}^*)^2 - (\vec{p}_{\mathrm{B}}^{\;*})^2} > 5.2$~\gevcc and $|\DeltaE=E_{\mathrm{B}}^*-E_{\mathrm{beam}}^*|<0.2$~GeV, where $\vec{p}_{\mathrm{B}}^{\;*}$ and $E_{\mathrm{B}}^*$ are, respectively, the momentum and energy of the reconstructed \B candidate in the CM frame, and $E_{\mathrm{beam}}^*$ is half the CM energy. The study is performed as a blind analysis, which means that the selection is optimized without examining the data in the signal region, \sigwind.

The PID efficiency for protons is larger than $99\%$ and the misidentification of kaons and pions as protons is less than $1\%$. The difference between the PID performance in data and simulation is evaluated using events from high-purity channels, which form the control samples (CS) for a given particle type.
For example, events with $\Lambda^0 \rightarrow p\pi^-$ decays form the CS for the validation of proton PID, $K_S^0 \rightarrow \pi^+\pi^-$ for pion PID, and $D^{*+} \rightarrow \pi^+D^0 ( D^0\rightarrow K^-\pi^+)$ for kaon PID. The PID efficiency of the MC-simulated events is corrected to match that observed in data by applying the weight $\epsilon_{\mathrm{CS, Data}}/\epsilon_{\mathrm{CS, MC}}$, where $\epsilon_{\mathrm{CS, Data}}$ and $\epsilon_{\mathrm{CS, MC}}$ are the PID efficiencies evaluated from the CS in data and simulation, respectively.

After applying the particle identification and preselection requirements, the fraction of misidentified signal candidates in simulation is found to be negligible ($<0.2\%$). The main background is combinatorial, from genuine protons in continuum ($e^+e^-\rightarrow q\bar{q}$) events. The continuum background is further reduced by imposing a signal-like selection on the output of a multivariate boosted decision tree (BDT) algorithm. The BDT classifier uses the following input variables: \deltae, $\cos\theta^*_B$, with $\theta_B^*$ the polar angle of the $B$~meson candidate with respect to the beam axis in the CM frame, and the event shape variables $R_2$ and $|\cos\theta_{\mathrm{TH}}|$, where $R_2$ is the ratio of the second to the zeroth Fox Wolfram moments~\cite{ref:FoxWolfram} and $\theta_{\mathrm{TH}}$ is the angle between the thrust axis~\cite{ref:thrust} of the $B$ candidate and that of the rest of the event in the \ups\ rest frame. These kinematic and topological variables are effective in discriminating between spherically shaped events from $B\bar{B}$ decays and jet-like $q\bar{q}$ events.

\begin{figure}
   \centering
   \includegraphics[scale=0.45]{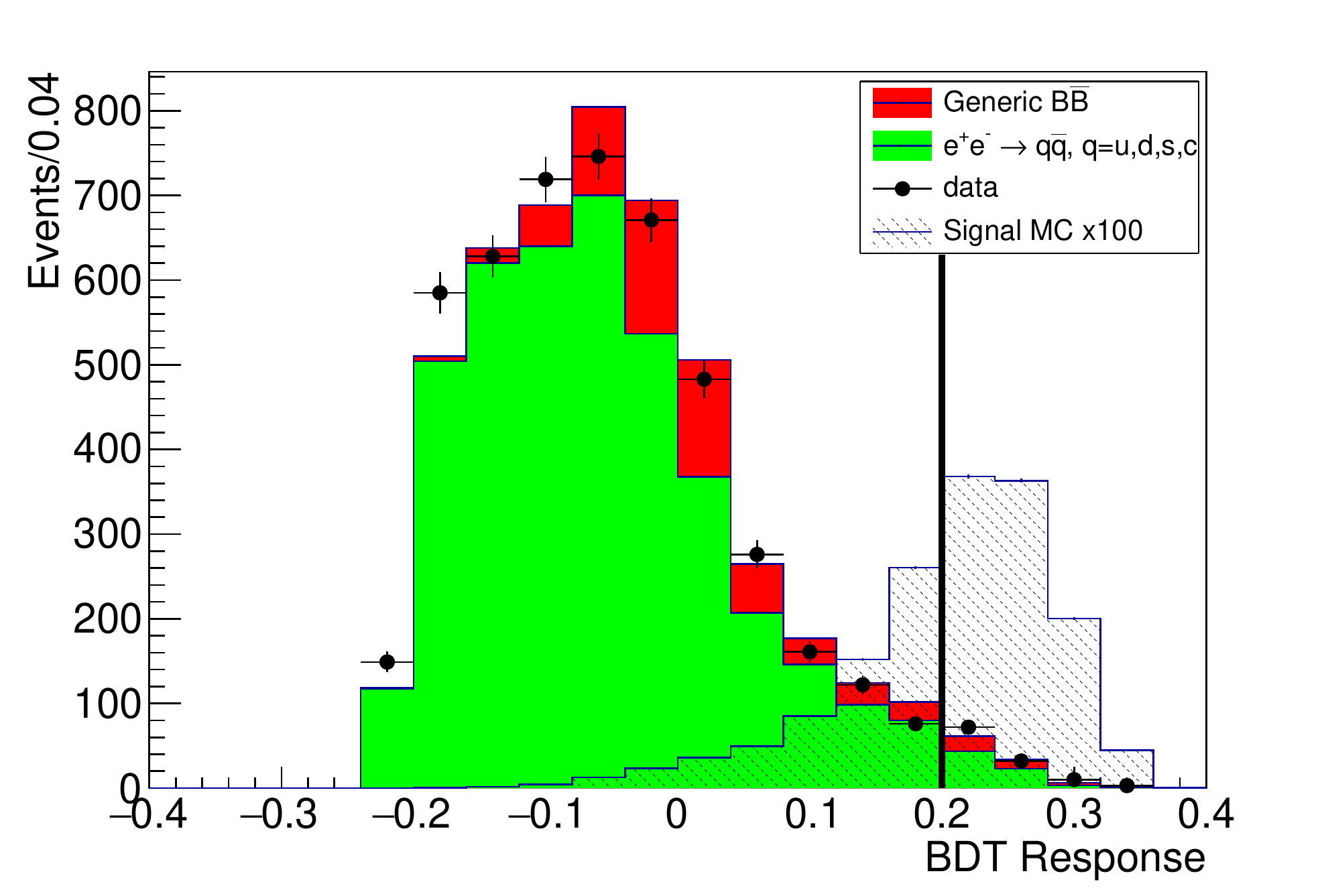}
   \caption{The BDT output distribution for simulated signal (shaded histogram) and background (filled histogram) events. The two background components, from continuum and generic $\B\Bb$ events, are stacked, with the total background prediction scaled to correspond to the number of selected events in the data. For purposes of visibility, the signal distribution has been multiplied by a factor of 100. The selection on the BDT output is indicated by the black vertical line.}\label{fig:bdtoutput}
 \end{figure}

\begin{figure}
  \centering
  \includegraphics[width=0.9\columnwidth, trim={0.15cm 0.15cm 0.65cm 0.9cm}, clip]{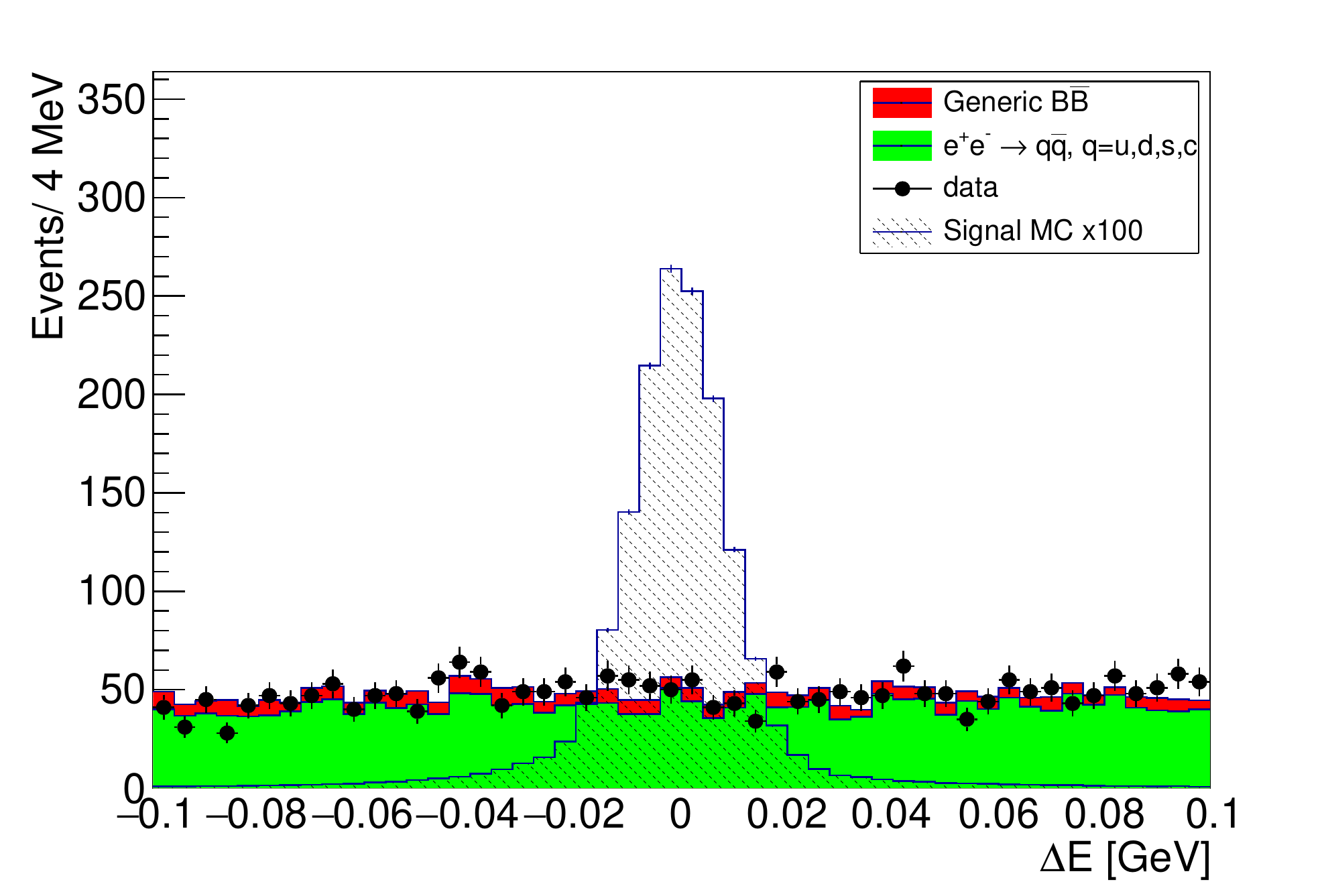}
  \includegraphics[width=0.9\columnwidth,trim= {0.15cm 0.15cm 0.65cm 0.9cm}, clip]{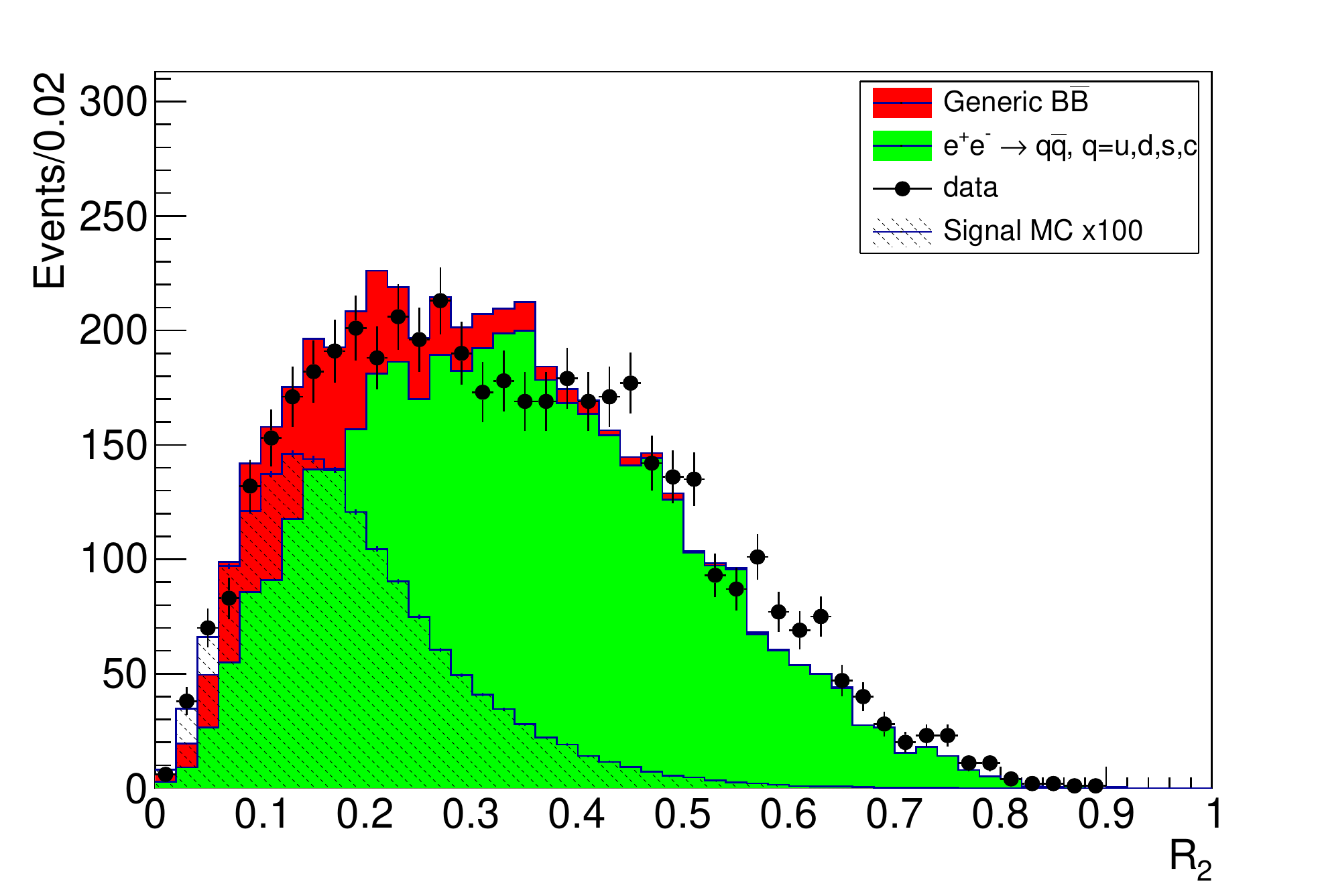}
  \includegraphics[width=0.9\columnwidth,trim={0.15cm 0.1cm 0.65cm 0.9cm}, clip]{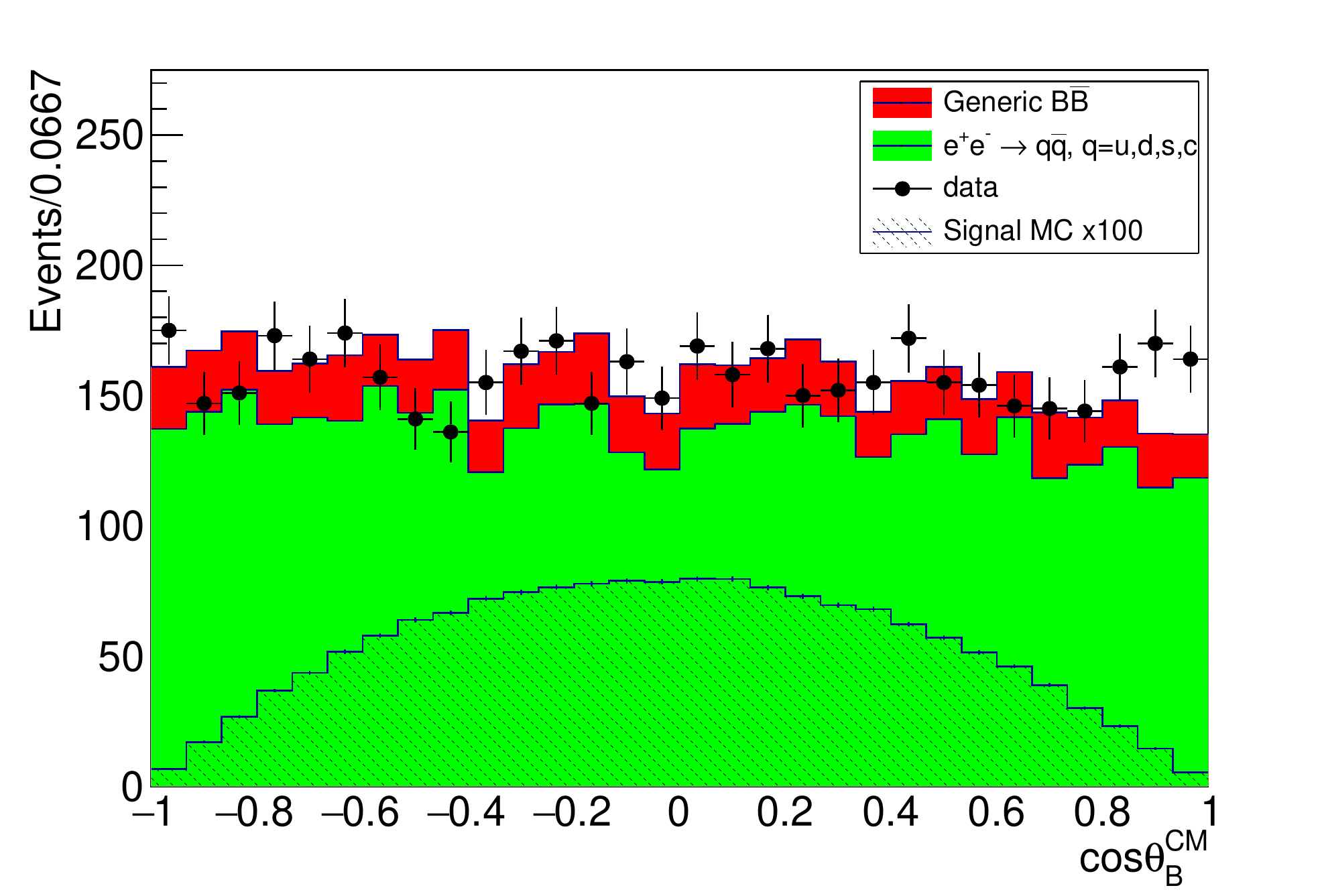}
  \includegraphics[width=0.9\columnwidth, trim={0.15cm 0.15cm 0.65cm 0.9cm}, clip]{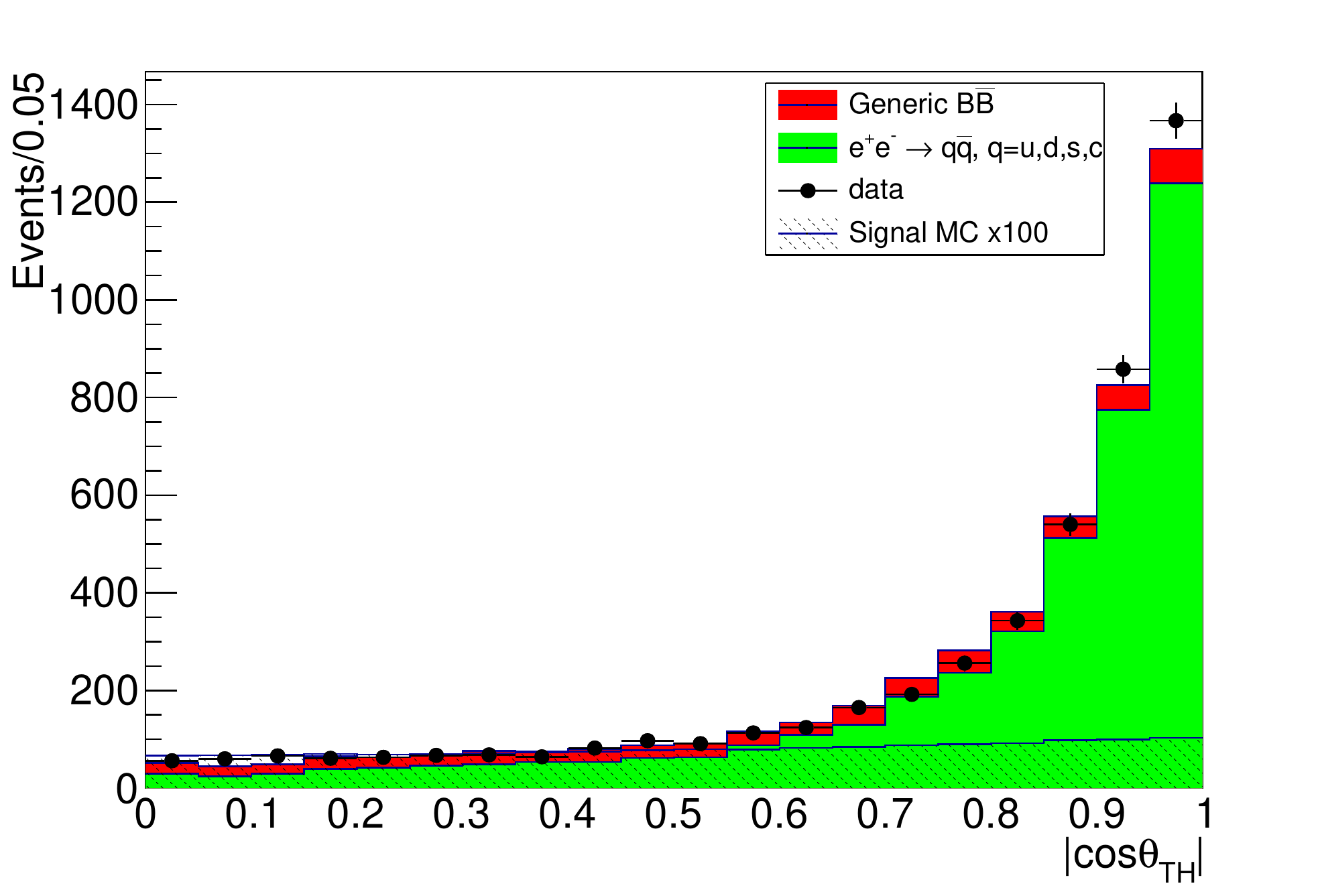}
  \caption{Comparison between data (dots) and MC (stacked histograms) for the BDT input variable distributions, before the BDT selection is applied. The total MC prediction for backgrounds has been scaled to correspond to the number of selected events in the data before the BDT selection, while the signal MC (shaded histogram) is multiplied by a factor of 100 for better visibility. }\label{fig:invars_beforeBDT}
    \end{figure}

 In the BDT output, signal (background) events peak at positive (negative) values (Figure~\ref{fig:bdtoutput}). The optimal selection on the BDT output is determined by maximizing the figure of merit \significance, where $S$ and $ B$ are the number of expected signal and background events, respectively. The number of signal events is estimated assuming the signal branching fraction of $10^{-7}$ mentioned above. 
The distributions of the input variables, before applying the BDT selection, are shown in Figure~\ref{fig:invars_beforeBDT}, where the signal and background MC samples have been normalized to match the number of selected events in data. The selection is optimized using the MC samples and is validated by comparing the distributions for the background MC samples to the data in the control region \sideband. Good MC--data agreement is observed.

The total number of selected data events in the interval \fitrange\ is $117$. The signal efficiency, evaluated from simulation, is found to be $\epsilon=0.2068\pm0.0004$ (stat).

%%%%BDT response
 We investigate the potential presence of peaking backgrounds from the baryonic modes $B\rightarrow p\bar{p} h^+h^-$, recently measured by the LHCb Collaboration~\cite{ref:LHCbBF}, where $h$ is a hadron other than a proton. These decays can potentially enter the background if the $h^+h^-$ pair is erroneously identified as a $p\bar{p}$ pair. This background is evaluated by applying the event selection to the simulated MC samples for the modes reported in Table~\ref{tab:peaking_bkg} and determing the selection efficiencies $\epsilon_{p\bar{p}h^+h^-}$. The number of expected background events in the data for each channel is estimated as $N_{p\bar{p}h^+h^-}= \epsilon_{p\bar{p}h^+h^-} \, \BR \, N_{B\bar{B}}$~(Table~\ref{tab:peaking_bkg}), where $\BR$ is the branching fraction measured in Ref.~\cite{ref:LHCbBF} and $N_{B\bar{B}}$ is the total number of $B\Bb$ pairs in the initial data sample. The expected contamination from these decays is found to be negligible.
\begin{table}
    \centering
  \caption{Potential peaking background modes. Branching fractions~\cite{ref:LHCbBF}, selection efficiencies, and the number of expected events at the data luminosity are reported.} \label{tab:peaking_bkg}
  \begin{tabular}{c c c c}
  \hline
  \hline 
  Decay & $\BR $ & Selection Efficiency & $N_{p\bar{p}h^+h^-}$  \\ 
mode  & $ (10^{-6})$ &  ($\epsilon_{p\bar{p}h^+h^-}$)& \\
\hline 
\pppipi & $ 2.7\pm0.4$& $(5\pm2)\times10^{-6}$ & $8.4\times 10^{-4}$ \\
\ppkk  & $0.11\pm0.03$ &  $(1.5\pm0.7)\times10^{-5}$& $7.8\times10^{-3}$\\
\ppkpi & $5.9\pm0.6$ & $(1.4\pm0.4)\times10^{-5}$ & $3.7\times10^{-2}$ \\
\hline
\hline
  \end{tabular}
    \end{table}
  To describe the \Mes\ distribution in data, we use a probability density function (PDF)  corresponding to the sum of the signal and the background components. The signal component is described by a Gaussian function, whose mean and width are fixed to values determined from a fit to simulated signal events. The combinatorial background component is described by the empirical ARGUS function~\cite{ref:argus_function}, which depends on two parameters: a shape parameter and a cutoff parameter. The cutoff parameter is set equal to the endpoint in the \Mes\ spectrum, $5.289$~\gevcc. The shape parameter is determined in the fit, along with the signal and background event yields, $N_{\mathrm{sig}}$ and $N_{\mathrm{bkg}}$, respectively.

\begin{figure}
    \centering
    \includegraphics[width=0.99\columnwidth]{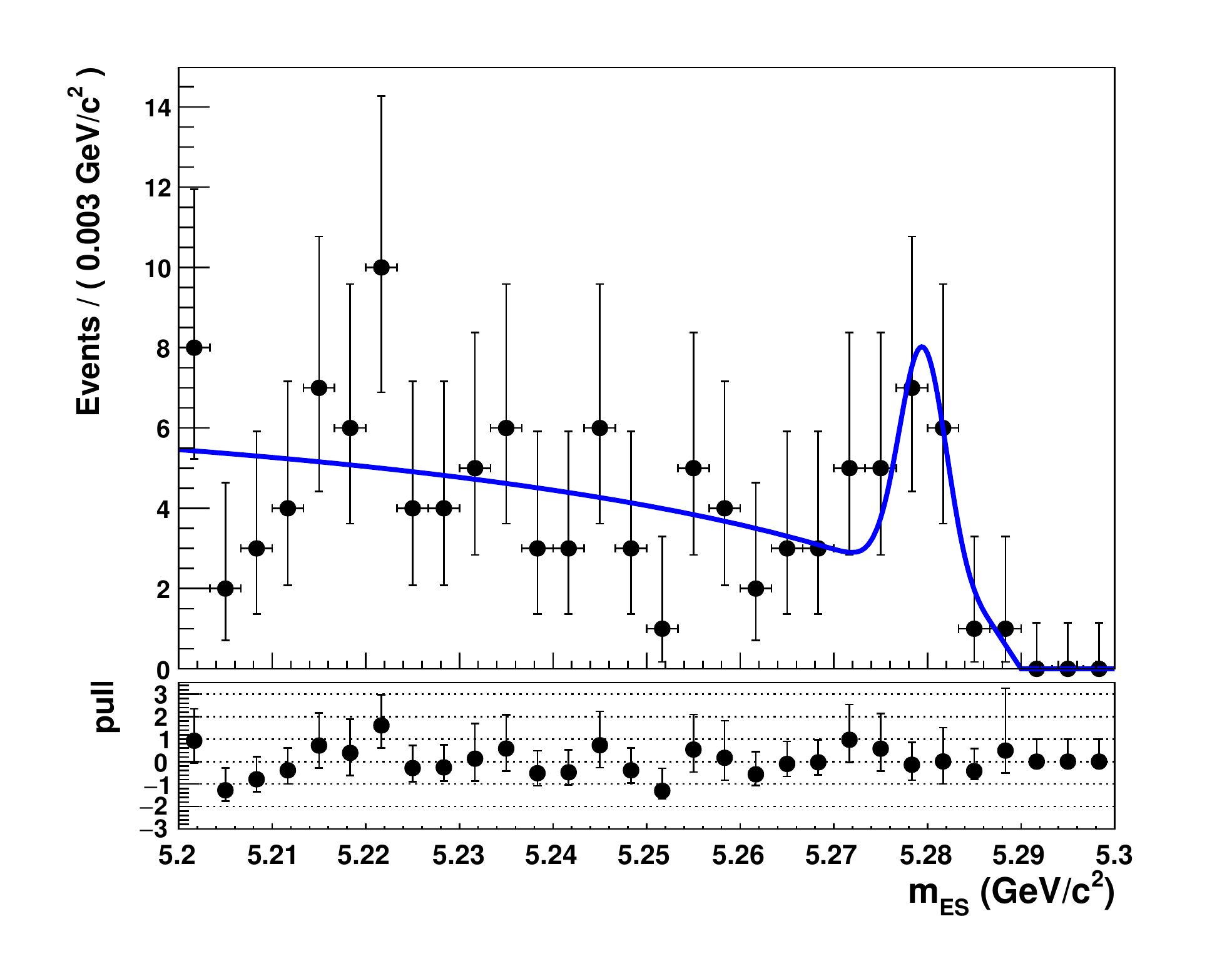}
    \caption{Fit to the data \Mes\ distribution (dots) in the interval \fitrange. The bottom plot shows the pull distribution, which is the bin-by-bin difference between the data and fitted distribution normalized by the corresponding statistical uncertainty from the fit.}\label{fig:unblind_fit}
  \end{figure}

  The signal yield is extracted by performing an unbinned extended maximum likelihood fit to the \Mes\ distribution in the range \fitrange~(Figure~\ref{fig:unblind_fit}). The logarithm of the extended likelihood is written as:

  \noindent
  \begin{align*}
  \log L(N_{\mathrm{sig}},N_{\mathrm{bkg}};x) &= - (N_{\mathrm{sig}}+N_{\mathrm{bkg}}) + \\
  & \sum^n_{i=1}\log(N_{\mathrm{sig}}\cdot f_{\mathrm{sig}}(x_i) + N_{\mathrm{bkg}}\cdot f_{\mathrm{bkg}}(x_i))
\end{align*}
\noindent
where $x$ corresponds to the measured \Mes\ distribution and the $f_j(x)$ are the corresponding PDFs for the signal and background components. The sum of the signal and background yields is constrained to the total number of observed events $n$.
  The result is $N_{\mathrm{sig}}=11.1 \pm 4.6$ (stat) events, from which the corresponding branching fraction is calculated as:
  
  \noindent
  \begin{align}\label{eq:bfr}
     \mathcal{B}(\text{\mymode})&=\frac{ N_{\mathrm{sig}}}{\epsilon \, 2\, N_{{B^{0}\Bzb}}}= \\ \nonumber
       &= (1.14 \pm 0.47)\times 10^{-7},
\end{align}
  \noindent

  where the uncertainty is statistical only and we assume equal production of $\Bz\Bzb$ and $B^+\B^-$ in \ups\ decays. Therefore, $2\,N_{{B^{0}\Bzb}} = N_{{B\Bb}}$, where \bpairs\ is the total number of $B\Bb$ pairs in the initial data sample. The value of 2 takes into account that the charge-conjugate decay is also reconstructed. The experimental values for $N_{\mathrm{sig}}$ and \bpairs\ are listed in Table~\ref{tab:input_value}.

To evaluate the statistical significance of the branching fraction result, we fit the data under a background-only hypothesis and determine the corresponding change $\sqrt{-2(\Delta\ln L)}$ with respect to the standard fit, where $L$ is the likelihood function. The statistical significance is found to be $2.9$ standard deviations. The systematic uncertainty in the ARGUS cutoff is taken into account and is found to not affect the signal significance.
\begin{table}
    \centering
    \caption{Experimental inputs used for the branching fraction calculation.
    } \label{tab:input_value}
  \begin{tabularx}{0.99\columnwidth}{c c c}
    \hline
    \hline
    Experimental Input & Value & Statistical uncertainty\\
   \hline 
   $N_{\mathrm{sig}}$ & $11.1$ & $4.6$\\
   $N_{B\Bb}$ & $470.88\times 10^{6}$ & $0.12\times 10^{6}$\\
    \hline  \hline
  \end{tabularx}
    \end{table}

Systematic uncertainties in the branching fraction arise from the uncertainty in \bpairs, from the fit procedure, and from the uncertainty in the signal efficiency. The relative systematic uncertainties for the considered sources are listed in Table~\ref{tab:list_of_sys}. The systematic uncertainty in \bpairs\ is estimated to be $0.6\%$~\cite{ref:bcounting}.

Potential systematic uncertainties associated with the fit procedure arise from the choice made for the signal PDF shape and from variation of the parameters held constant in the fit.  Variations of the form chosen to model the signal PDF are found to have a negligible impact on the result, while the uncertainty associated with the ARGUS cutoff value, evaluated by varying the cutoff value within its uncertainty of 0.5\mevcc, is 0.9\%

To determine the systematic uncertainty in the signal efficiency, several sources are taken into account: the statistical uncertainty from the MC samples, the PID performance, the track finding efficiency, the BDT method, and the decay model used for the generation of the signal MC sample. The finite size of the signal MC sample results in a relative systematic uncertainty of $0.2\%$. The PID performance contribution is taken as the effect of the full data-to-MC correction mentioned above and corresponds to a relative uncertainty of $0.9\%$. The systematic uncertainty related to the track finding efficiency is a function of the particle momentum~\cite{ref:tracking} and amounts to $0.9\%$ for protons of approximately $1\gevc$ momentum~\cite{ref:tracking}. The systematic uncertainty in the signal efficiency introduced by the BDT method is evaluated by reweighting, separately for each of the four input variables of the BDT classifier, the shape of the MC distribution to match that observed in data. The weights are calculated in the control region \sideband, before the BDT selection, as the bin-by-bin ratio between data and MC events, and are applied to the corresponding distribution of the signal MC sample. The difference in the efficiency computed with and without the weights applied provides a systematic uncertainty of $2.2\%$.

The systematic uncertainty related to the unknown dynamics of \mymode\ decays is evaluated by comparing the pure phase space decay MC sample to a model in which the decay proceeds through an intermediate spinless resonance,  $B \rightarrow X(\rightarrow p\bar{p})X(\rightarrow p\bar{p})$. Weights, binned in the 4-dimensional space of the magnitudes of the momenta of the four tracks, are obtained by dividing the momentum distribution resulting from the resonant model by that from the phase space model, and are applied to the proton momentum distribution of the signal MC. The systematic uncertainty is obtained from the difference in the efficiency computed from the weighted and unweighted samples. The largest relative difference is obtained for an $X$ mass of $2.6 \gevcc$. It amounts to $14\%$ and is the largest contribution to the total systematic uncertainty.

 The final result for the branching fraction is $\mathcal{B}(\text{\mymode}) = (1.14 \pm 0.47_{\mathrm{stat}} \pm 0.17_{\mathrm{sys}}) \times 10^{-7}$.
\begin{table}
    \centering
    \caption{Relative systematic uncertainties in the signal branching fraction. The total systematic uncertainty is determined by summing the individual contributions in quadrature.} \label{tab:list_of_sys}
\begin{tabularx}{0.99\columnwidth}{c c c}
  \hline
  \hline
  Variable & Source & Relative systematic\\
  && uncertainty (\%)\\
    \hline
   $N_{B\Bb}$ & $B$ counting & $0.6$\\
   $N_{\mathrm{sig}}$ & ARGUS cutoff & $0.9$\\
    $\epsilon$ & MC statistics& $0.2$ \\
    $\epsilon$ & PID efficiency & $0.9$ \\
    $\epsilon$ & Track finding efficiency & $0.9$ \\
    $\epsilon$ & BDT selection& $2.2$ \\
   $\epsilon$ & Decay model & $14$\\
   \hline
   Total & & $15$ \\
   \hline
   \hline
  \end{tabularx}
    \end{table}

The upper limit at $90\%$~C.L. on the signal yield is computed by integrating the likelihood as a function of $N_{\mathrm{sig}}$, up to the value $N_{\mathrm{sig}}^{\mathrm{UL}}$ such that the equality $\int_0^{N_{\mathrm{sig}}^{\mathrm{UL}}} L(N_{\mathrm{sig}}) dN_{\mathrm{sig}} =  0.90\int_0^{+\infty}L(N_{\mathrm{sig}}) dN_{\mathrm{sig}}$ is satisfied. This calculation is based on the Bayesian approach, assuming a uniform prior for $N_{\mathrm{sig}} > 0$ and $0$ otherwise. The $90\%$ C.L. upper limit on $N_{\mathrm{sig}}$ is $N_{\mathrm{sig}}^{\mathrm{UL}}=19$ events, corresponding to a 90\%~C.L. upper limit on the signal branching fraction $2.0 \times 10^{-7}$.

We use pseudoexperiments to establish the robustness of our result for the upper limit against systematic variation. The signal yield $N_{\mathrm{sig}}$ is varied according to the signal PDF computed from the fitted likelihood function and the signal efficiency $\epsilon$ is randomly smeared according to a Gaussian distribution with mean $0.2068$ and with width $0.03$, corresponding to its absolute systematic uncertainty. For each pseudoexperiment, the branching fraction is calculated from the input values for $N_{\mathrm{sig}}$ and $\epsilon$ randomly selected from the above defined PDFs and its distribution is integrated up to $90\%$~C.L., which shows a smeared upper limit consistent within its uncertainty with the unsmeared result.

In summary, we have performed a search for $B$ meson decays to the $pp\bar{p}\bar{p}$ final state, obtaining $11.1$ signal events. The statistical significance of the result is 2.9 standard deviations. The branching fraction is measured to be ${\cal{B}}=(1.1\pm0.5\text{(stat)}\pm0.2\text{(syst)})\times 10^{-7}$. The corresponding 90\% C.L. upper limit is $\mathcal{B}(\text{\mymode}) < 2.0 \times 10^{-7}$.
Our result can provide important input for QCD models of hadronization and improve understanding of the threshold enhancement effect.

\input acknowledgements

\end{document}

%% file: authors_mar2018_frozen.tex
% NOTES
% 10-Mar-2018 Add E. A. Kozrev to author list                          J.W. Gary
% 21-Feb-2018 Add footnote for Walter Innes as ``deceased''            J.W. Gary
% 23-Nov-2017 Move Hossain Ahmed from Jazan to St. Francis Xavier      J.W. Gary
%    and adjust list of BaBar institutes on M. Roney's instructions
% 15-Oct-2017 Add three new authors to frozen author list:             J.W. Gary
%     Laura Zani, Yunxuan Li, Robert Seddon on M. Roney's instructions
% 04-Oct-2017 Correct the postal code for Wuhan U.                     J.W. Gary
% 22-Jan-2017 Changes requested by Babar Spokesperson (M. Roney)       J.W. Gary
%    Remove V. Luth; add H. Lacker and R. Sobie
%    Add Institute of Particle Physics label for C. Hearty, S. Robertson,
%    R. Sobie and ``a'' ``b'' superscripts for all Canadian authors
%    Move R. Cheaib from McGill to Mississippi
% 20-Jan-2017 Add ``deceased'' footnote for Erwin Gabathuler           J.W. Gary
% 16-Dec-2016 Move Miriam Fritsch from Mainz to Bochum                 J.W. Gary
% 04-AUG-2016 Add ``deceased'' footnote for Giancarlo Piredda          J.W. Gary
% 02-JUN-2016 Move Marcello Rotondo from Padova to Frascati            J.W. Gary
% 20-FEB-2016 Add footnote for Liang Sun                               J.W. Gary
% 21-DEC-2015 Add Bologna alternative address for Claudia Patrignani   J.W. Gary
% 
\author{J.~P.~Lees}
\author{V.~Poireau}
\author{V.~Tisserand}
\affiliation{Laboratoire d'Annecy-le-Vieux de Physique des Particules (LAPP), Universit\'e de Savoie, CNRS/IN2P3,  F-74941 Annecy-Le-Vieux, France}
\author{E.~Grauges}
\affiliation{Universitat de Barcelona, Facultat de Fisica, Departament ECM, E-08028 Barcelona, Spain }
\author{A.~Palano}
\affiliation{INFN Sezione di Bari and Dipartimento di Fisica, Universit\`a di Bari, I-70126 Bari, Italy }
\author{G.~Eigen}
%\author{B.~Stugu}
\affiliation{University of Bergen, Institute of Physics, N-5007 Bergen, Norway }
\author{D.~N.~Brown}
%\author{L.~T.~Kerth}
\author{Yu.~G.~Kolomensky}
%\author{M.~J.~Lee}
%\author{G.~Lynch}
\affiliation{Lawrence Berkeley National Laboratory and University of California, Berkeley, California 94720, USA }
\author{M.~Fritsch}
\author{H.~Koch}
\author{T.~Schroeder}
\affiliation{Ruhr Universit\"at Bochum, Institut f\"ur Experimentalphysik 1, D-44780 Bochum, Germany }
% ...
\collaboration{The \babar\ Collaboration}
\noaffiliation

%% file: acknowledgements.tex
We are grateful for the 
extraordinary contributions of our \pep2\ colleagues in
achieving the excellent luminosity and machine conditions
that have made this work possible.
The success of this project also relies critically on the 
expertise and dedication of the computing organizations that 
support \babar.
The collaborating institutions wish to thank 
SLAC for its support and the kind hospitality extended to them. 
This work is supported by the
US Department of Energy
and National Science Foundation, the
Natural Sciences and Engineering Research Council (Canada),
the Commissariat \`a l'Energie Atomique and
Institut National de Physique Nucl\'eaire et de Physique des Particules
(France), the
Bundesministerium f\"ur Bildung und Forschung and
Deutsche Forschungsgemeinschaft
(Germany), the
Istituto Nazionale di Fisica Nucleare (Italy),
the Foundation for Fundamental Research on Matter (The Netherlands),
the Research Council of Norway, the
Ministry of Education and Science of the Russian Federation, 
Ministerio de Econom\'{\i}a y Competitividad (Spain), the
Science and Technology Facilities Council (United Kingdom),
and the Binational Science Foundation (U.S.-Israel).
Individuals have received support from 
the Marie-Curie IEF program (European Union) and the A. P. Sloan Foundation (USA). 

% NOTES:
% add "and the Binational Science Foundation (U.S.-Israel)"  07-Oct-2013 Bill Gary (Abi Soffer request)